# CONCURRENT SOFTWARE DESIGN
# BASED ON CONSTRAINTS ON STATE DIAGRAMS[1]

Bogdan D. Czejdo

Department of Mathematics and Computer Science

Loyola University 70118

New Orleans, LA

Email: czejdo@loyno.edu

Wiktor B. Daszczuk, Jerzy Miescicki

Warsaw University of Technology

Institute of Computer Science

Nowowiejska 15/19, 00-665 Warsaw, Poland

Email: wbd@ii.pw.edu.pl, jms@ii.pw.edu.pl

## ABSTRACT

Concurrent software for engineering computations consists of multiple cooperating modules. The behavior of individual modules is described by means on state diagrams. In the paper, the constraints on state diagrams are proposed, allowing for the specification of designer's intentions as to the synchronization of modules. Also, the translation of state diagrams (with enforcement constraints) into Concurrent State Machines is shown, which provides formal framework for the verification of inter-module synchronization. An example of engineering software design based on the method is presented.

## 1. INTRODUCTION

Modern computational systems very often include multiple modules cooperating with each other. Such systems typically require sophisticated concurrent software to allow for proper synchronization of cooperating modules. In [Basz95, Basz96a, Basz96b] we have described how to create and verify concurrent software for engineering design. We used the approach based on state diagrams which allow for explicit and visual specifications of interactions

---

[1] This work was supported by Grant N 8T11C00708 from Polish State Commitee for Scientific Research

between software modules. The state diagrams were converted into Concurrent State Machines [Mie92a, Mie92b, Mie94] and their reachability graph was derived, allowing for the analysis of the concurrent behavior of modules.

However, as shown in [Mie96], the synchronization of concurrent diagrams is a nontrivial issue and complete specification requires rather specialized knowledge. That could be too much burden for an average engineer. In this paper we will use the constraints on the state diagram to describe only the intentions of the engineer. The appropriate synchronization can be constructed automatically from these constraints. We can identify at least three basic types of constraints: enforcement, exclusion and preemption. In this paper we will discuss the enforcement constraints that allow to specify the synchronization of modules when state changes in one module **force** some state changes in another module.

The paper is organized as follows. Section 2 describes briefly the state diagram method for software specification and introduces enforcement constraints themselves. The conversion from constraints to concurrent automata (CSM) models is discussed in Section 3. An example of engineering software design based on our method is shown in Section 4.

## 2. ENFORCEMENT CONSTRAINTS ON STATE DIAGRAMS

We assume that the description of all concurrent modules is done using a behavioral model [Emb92, Rum91, Shl88, You89]. This model is an extended state net that describes the dynamic behavior of objects. It has three basic components: *states* (for each object in the given class), *triggers* that cause the transition of an object from one state to another and *actions* performed during the transition. Generally triggers can be either Boolean conditions or events. In this paper, however, we will not use Boolean conditions. We will convert each Boolean condition into an appropriate event.

In state diagrams, each state is represented by a rounded box while each transition is indicated by labeled arrow. The first part of label (before the slash) specifies the trigger and the other part (after the slash) specifies the action to be performed during transition. The actions can be just simple events sent to other models. They can also involve some complex activities that consist of several steps.

A simple model of a concurrent software design situation is shown in Figure 1. The system under consideration consists of two modules. Each module is designed separately and the designer does not need to be concerned with the interactions of these two modules. The left-hand module (or L_Module) includes two states: *1* and *2*, and the right-hand module (or R_Module) contains two states: 3 and 4. The transitions between states *1* and *2* are triggered by external events *z12* and *z21*. Similarly, the transitions between states *3* and *4* are triggered by external events *z34* and *z43*. In general there may be more transitions between the states but they will be treated identically in our approach and therefore can be omitted. Transitions might involve some operations, but such operations are irrelevant to our transformations and therefore they will not be considered here.

Specification of enforcement constraints requires from the designer to identify some elements of the system, at least:

- the modules to be synchronized (perhaps not all modules),
- the states that have to be in constraints relation with states of other modules.

Generally, the process of specifying the constraints requires to identify the states (or super-states) in different modules that are involved in the constraints. At this stage, we do not care about transitions. The purpose is to identify the subjects of constraints (selected modules and their states) and the type of constraints. Of course, there may be also many other dependencies between modules, but our target is to specify synchronization constraints only. Such specification lets us express the semantic dependencies, which are difficult to express in state diagrams themselves. To express constraints, some authors propose a variety of shapes and styles of arrows representing various transitions and dependencies between states. Our approach is to describe constraints on higher level of abstraction, where there is no concern for transitions and only semantic constraints on synchronization are being specified.

There are four possible enforcement constraints specified graphically in Figure 1: enforcement L→R (*1,3*), enforcement L→R (*2,4*), enforcement R→L (*3,1*), and enforcement R→L (*4,2*). They describe the request for the following synchronization. If the state of L_Module is changed to *1* then the state of R_Module must become *3*. Similarly, if the state of L_Module is changed to *2*, the state of R_Module has to become *4*, etc. The intentions of the designer represented by enforcement constraints can be translated automatically into explicit diagram interactions. Below, we discuss several cases of these interactions.

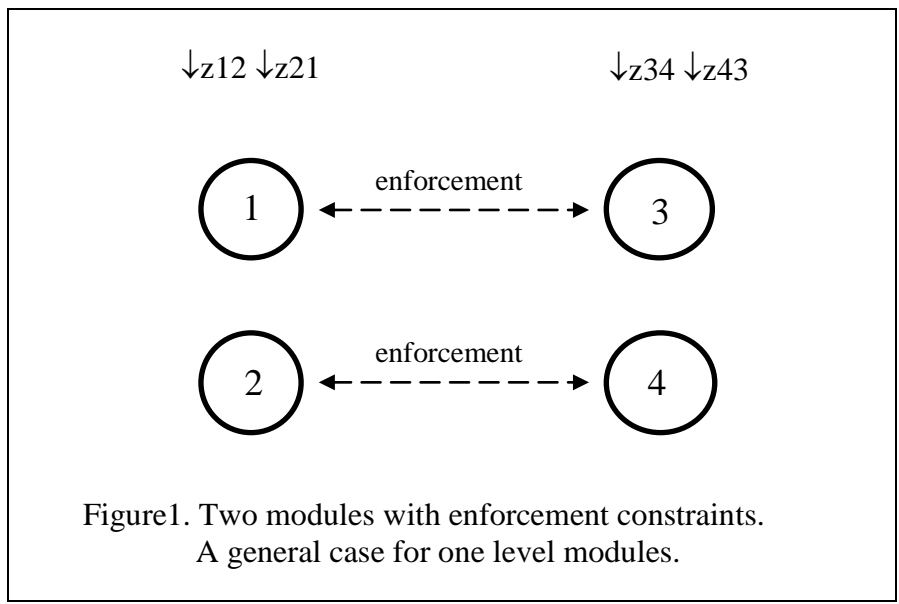

Figure1. Two modules with enforcement constraints.
A general case for one level modules.

First, let us consider the system consisting of two modules (L and R) with no external events coming to R_Module. In this case the synchronization intentions can be shown in Figure 2 (constraints expressed in upper-right corner of Fig. 2). L_Module has to transmit two events *s34* and *s43* while R_Module has two transitions trigged by *s34* and *s43*. This is an example of a simple master-slave synchronization.

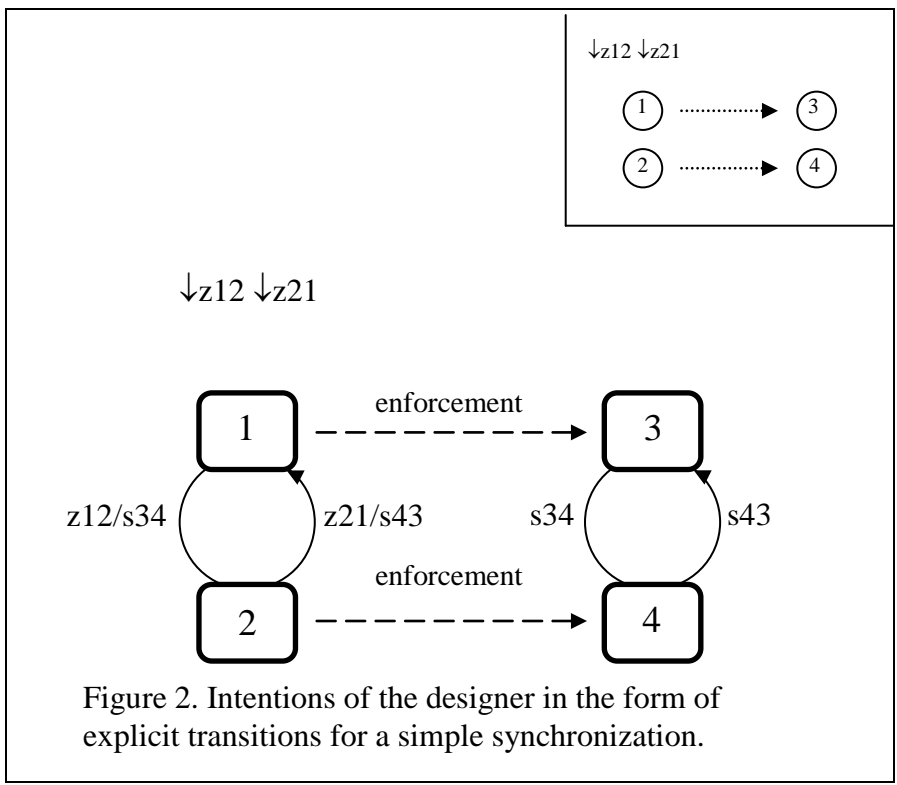

Figure 2. Intentions of the designer in the form of explicit transitions for a simple synchronization.

Next, let us consider the system consisting of two modules with external events coming to both modules but the R_Module receives only the signal *z43*. In this case the synchronization intentions can be shown in Figure 3. L_Module has two events *s34* and *s43* and one transition on *s21*, while R_Module has two transitions on *s34* and *s43* and one event *s21*.

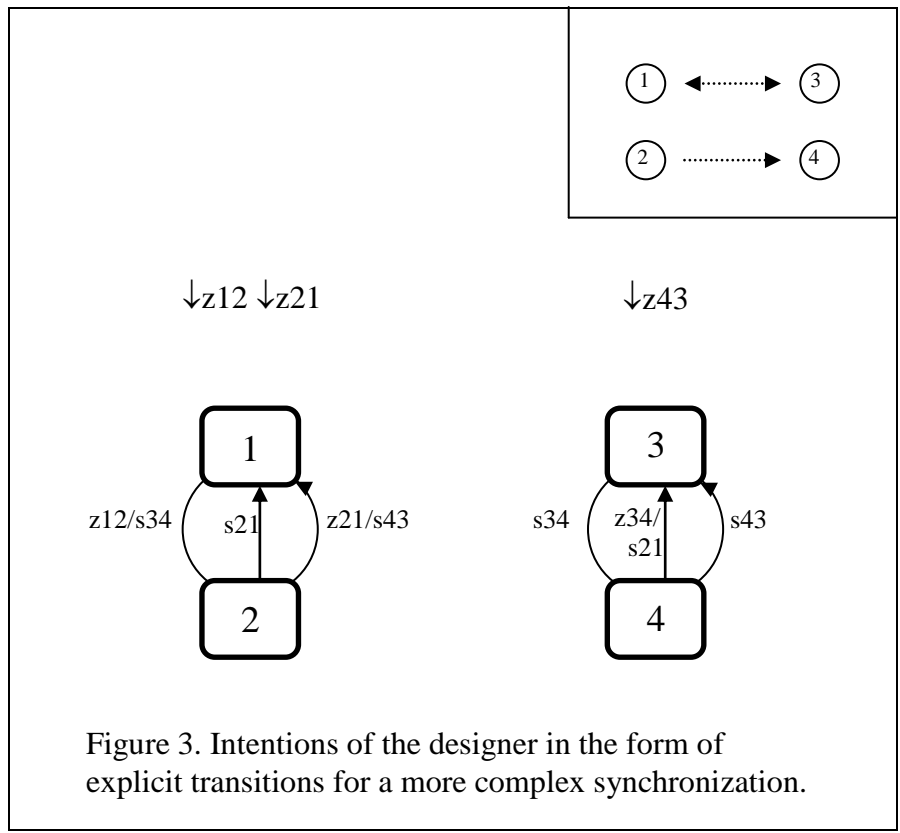

Figure 3. Intentions of the designer in the form of explicit transitions for a more complex synchronization.

In general, the state diagram can be nested or multi-level one. There are many interesting problems related to the multi-level diagrams [Basz96b, Rum91]. One of the fundamental problems is identification of initial state(s) of the lower level diagram and states from which we can exit from the lower level diagram. The problems related with the initial state(s) are relatively well described in [Rum91, Har87, Har88]. The states from which we can exit can be described in terms of final transitions [Emb92]. We will extend this approach by using hierarchy of events.

Let us consider the system with R_Module represented by a multi-level diagram as shown in Fig. 4. The constraints are identical to those presented in upper-right corner of Fig. 3. The abstract (high level) state 4 is described by a sub-diagram containing two states: $4_1$ and $4_2$.

___

Moreover, we specify one more level of hierarchy of events. The event *z43* from Fig. 3 is split to a pair of events: *x43* and *y43*.

The hierarchy of events is presented in Fig. 5. In second row, Env→L denotes external signal addressed to L_Module, while Env→R – to R_Module. Tehre are also inter-module events: generated by the L_Module and send to the R_Module (L→R) and generated by the R_Module and send to the L_Module (R→L). Third row enumerates the signals: external ($z_{ij}$) and inter-module for implementation of constraints ($s_{ij}$). The pair of events *x43* and *y43*, constituting together the abstract event *z43* is presented in fourth row. Last row shows the implementation of events *x43* and *y43* as $x4_13$, $y4_13$ and $y4_23$ inside refined state *4*.

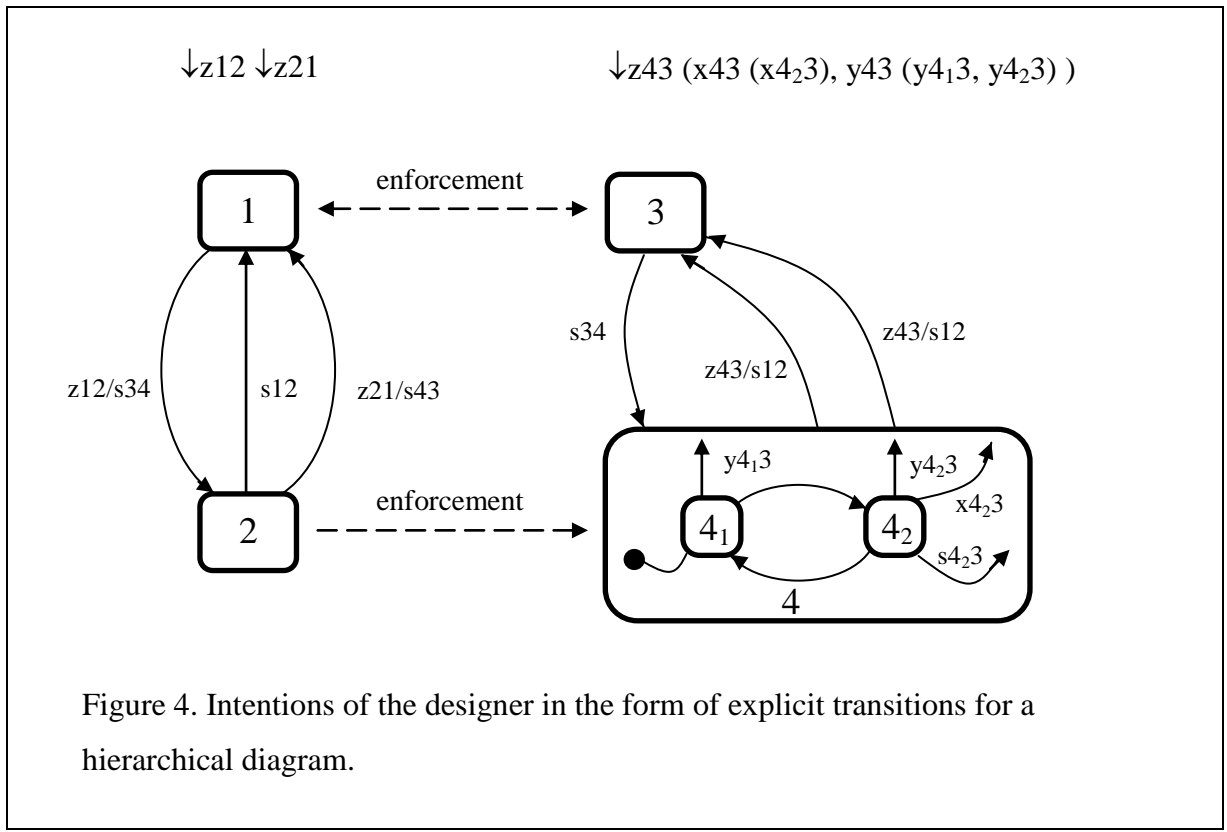

Figure 4. Intentions of the designer in the form of explicit transitions for a hierarchical diagram.

The signals *x43* and *y43* have identical consequences (transition from *2* to *1* in L_Module), but they have different meanings. They have also different implementations: the event *x43* will be accepted only in state $4_2$, while the event *y43* will be accepted in both states $4_1$ and $4_2$, but we define it as two separate events: $y4_13$ accepted only in state $4_1$ and $y4_23$ accepted only in state $4_2$. Inter-module signal *s43* will be accepted only in state $4_2$ as $s4_23$. Note that the hierarchy of events is a design decision.

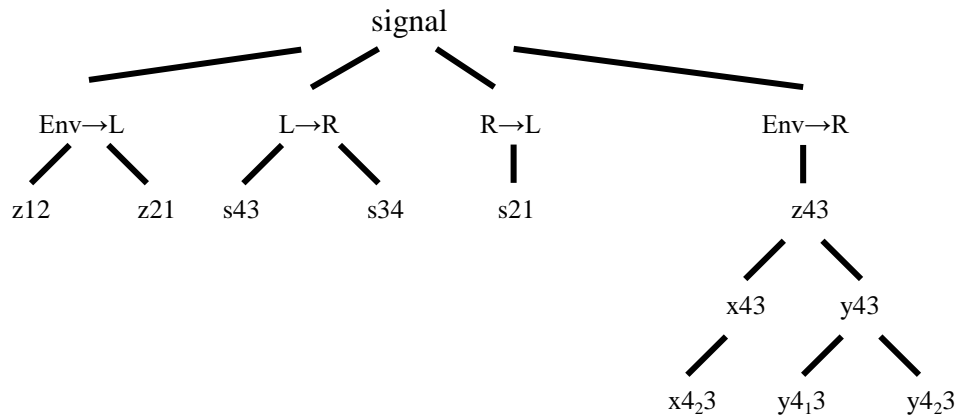

Figure 5. Classification of external and inter-module events.

## 3. TRANSFORMATION OF ENFORCEMENT CONSTRAINTS ON STATE DIAGRAMS INTO CSM MODELS

State diagrams are used to describe the ordering of events, actions performed in states, signals sent and received by modules etc. However, if we deal with concurrent system components - state diagrams themselves do not provide enough information for a formal verification of system synchronization. We may find out or otherwise identify e.g. the possible system deadlock or the violation of mutual exclusion but we cannot prove (from state diagrams themselves) that the system is deadlock-free, live etc. To verify the system's synchronization rules formally we will apply an approach based on Concurrent State Machines (CSM, [Dasz92, Dasz93, Kol92, Mie92a, Mie92b, Mie94]).

Concurrent State Machines can be viewed as graphs, resembling the known and conventional finite state machines (FSM), consisting of a finite set of nodes (states) connected by labeled arrows (transitions). However, while in conventional FSM the transitions are labeled with single symbols from the input alphabet, in CSM the labels are Boolean formulas interpreted in a following way.

Let, for instance, the transition from $i$ to $j$ be labeled by the formula '$\sim a*b$'. This means that the transition is executed whenever the machine (in state $i$) receives any combination of symbols that does not contain $a$ and contains $b$. Single '$b$' (or one-element set $\{b\}$) is surely just such an input but also an infinite number of other coincident occurrences of symbols (e.g. $\{b, c\}$, $\{b, d\}$, $\{b, c, d\}$, ... ) fulfill this condition. This way, in CSM we can specify the required machine's reaction on single symbols as well as on coincident occurrence of symbols from the arbitrarily large alphabet. If (for a given state and a given input) two or more Boolean formulas are true then a non-deterministic choice is made and only one out of 'enabled' transitions is actually executed. Note also that the Boolean formula **1** means 'always' (or 'unconditionally') while Boolean **0** means 'never'. The transitions labeled with **0** are never executed, thus we can simply erase them from the graph.

CSM can also produce (or transmit) similar sets its output symbols. These outputs are produced by CSM states (as in Moore-type finite automata), not by transitions. Moreover, it is assumed that in a system of several machines there is always some communication medium which broadcasts (immediately and faultlessly) to all system components the union of all sets of symbols produced by the system environment and components themselves. Thus, given a joint system state (i.e. a vector of states of individual system components), we know exactly what is the joint output 'audible' in this state to all machines. We can also compute (using Boolean product) the Boolean labels on transitions from this particular state to other system states. Of course, only non-**0** labels signify actually possible transitions. They lead to states reachable from a given one. Starting this procedure from the system's initial state we can derive the graph of system's reachable states which makes a full and formal picture of system's behavior.

In the Institute of Computer Science, Warsaw University of Technology, a software tool named COSMA has been developed for the computation and analysis of reachability graphs of systems of CSMs. The examples discussed below were solved with the use of COSMA 1.0. The next version of this software, supporting the analysis of systems of nested CSMs, is now under development.

The translation of state diagrams into CSM can be illustrated as in Fig. 6b. Firstly, the state diagrams produce their outputs on transitions while CSM do it in states. Thus, rewriting state diagrams into CSM one has to split each output-producing transition into two steps, separated by the additional state which produces just this particular output. Secondly, we assumed that the component which receives a message from the partner (e.g. R_Module receiving *s34* from L_Module) must acknowledge it by sending the appropriate *ac_xxx* symbol (e.g. *ac_s34*), while the machine which sends a message (e.g. L_Module sending this *s34*) must wait for the acknowledgement before proceeding to further activities. Symbols from the environment (*z12* or *z21*, for instance) do not need to be confirmed this way. This additional mechanism (which is not explicitly required in state diagrams) was introduced in order to make the inter-module communication more realistic. Moreover, while our simple system of two two-state machines would behave properly without confirmations, it can be shown that in a more general case the lack of acknowledgements can easily result in a severe synchronization fault, e.g. deadlock or livelock.

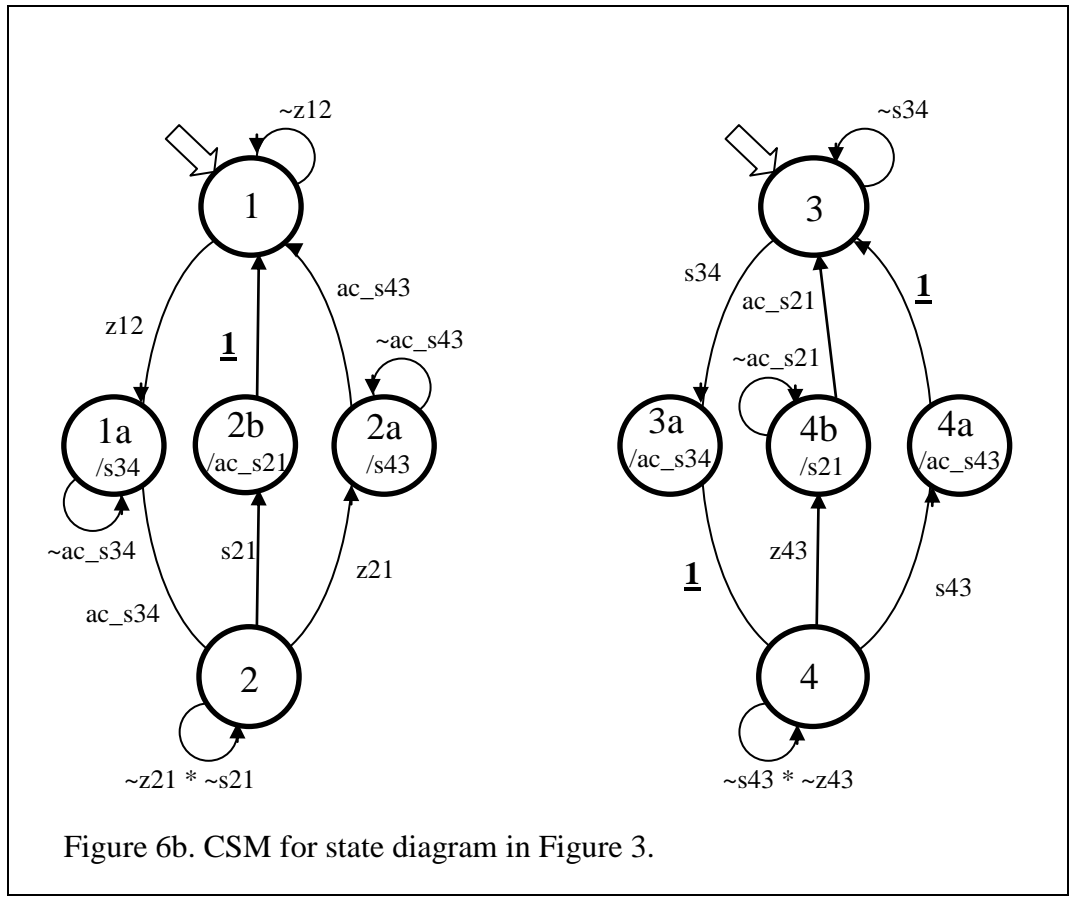

Figure 6b. CSM for state diagram in Figure 3.

The reader is encouraged to check how the superposition of the two above-sketched rules can be seen in Fig. 6b, which shows the transformation of the state diagram from Fig. 3 into a system of CSM. The system's reachability graph (Fig. 6c) reveals a deadlock in the state (*2a, 4b*) which is not so easily seen from Fig. 6b itself. System's reachability graph provides the clues for a more detailed analysis of component's behavior. Indeed, the deadlock can actually occur either upon the exact coincidence of two signals from the environment (*z21* and *z43*) when the system is in (*2, 4*) or when the other of these two signals comes at 'too early stage' of the reaction on the first one.

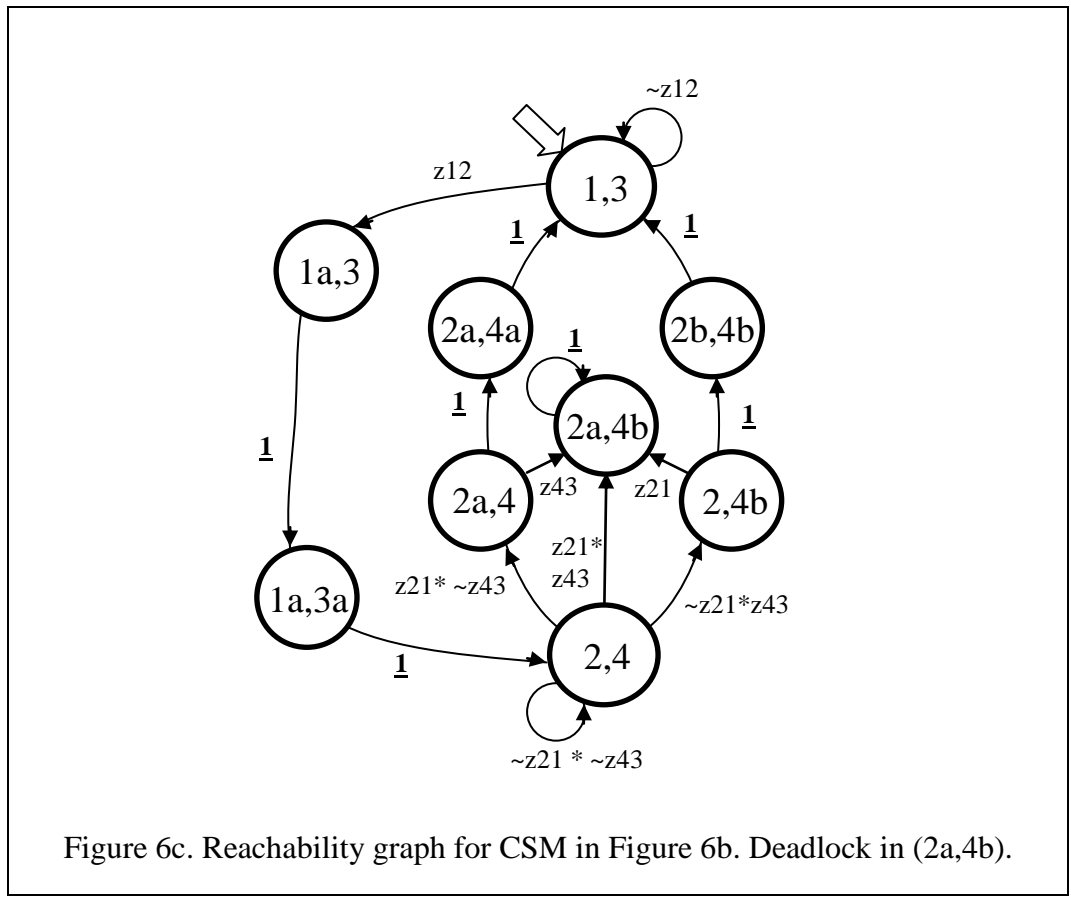

Figure 6c. Reachability graph for CSM in Figure 6b. Deadlock in (2a,4b).

The identification of deadlock causes allows us to modify both modules as shown in Fig. 6d. This case, the reachability graph (Fig. 6d) contains no deadlock, however the remedy applied (producing additional, 'redundant' acknowledgements in states 2a and 4a) cannot be viewed as a universal or unconditionally recommended solution.

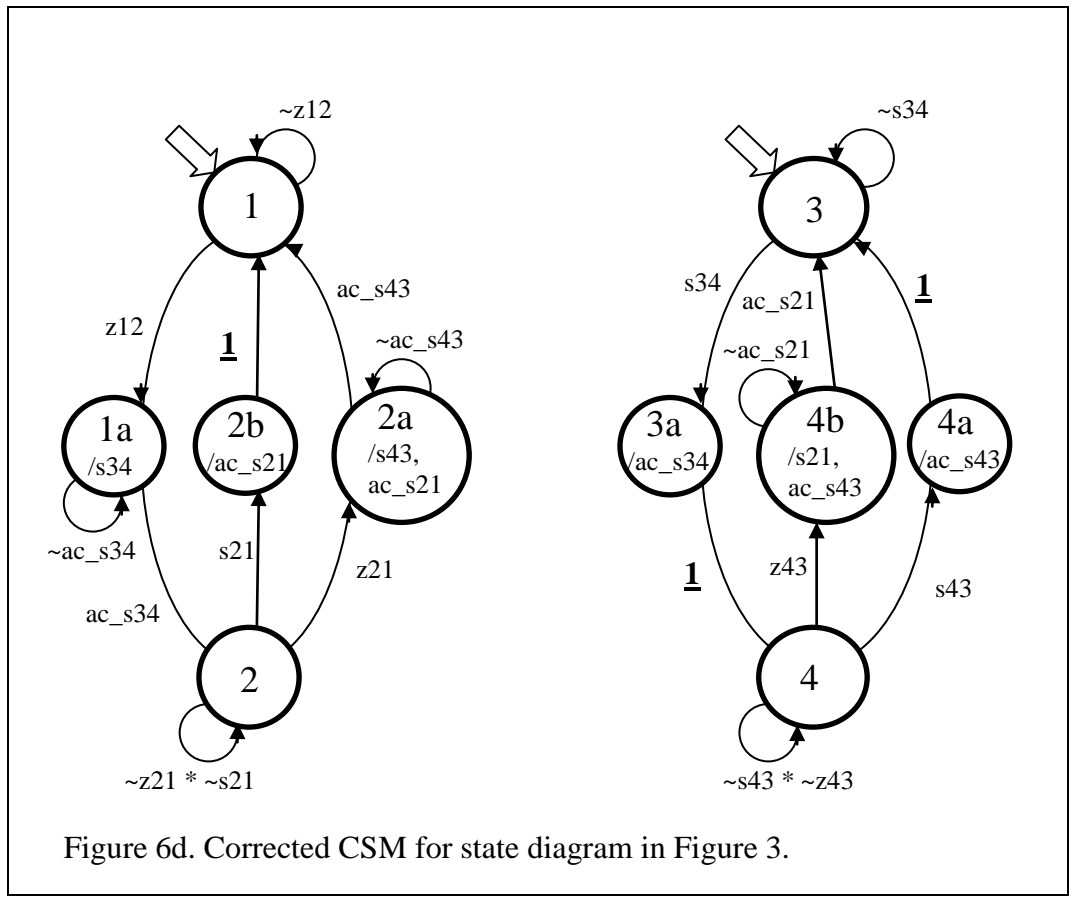

Figure 6d. Corrected CSM for state diagram in Figure 3.

As it was mentioned above, the nested CSM can be also used and the appropriate software (COSMA 2.0) for their analysis is now under development. The example CSM model, corresponding to state diagrams from Fig. 4, is shown in Fig. 6f. For simplicity, Boolean labels of transitions that do not go outside of the super-state *4* (leading to states $4_1$ and $4_2$) are not shown here.

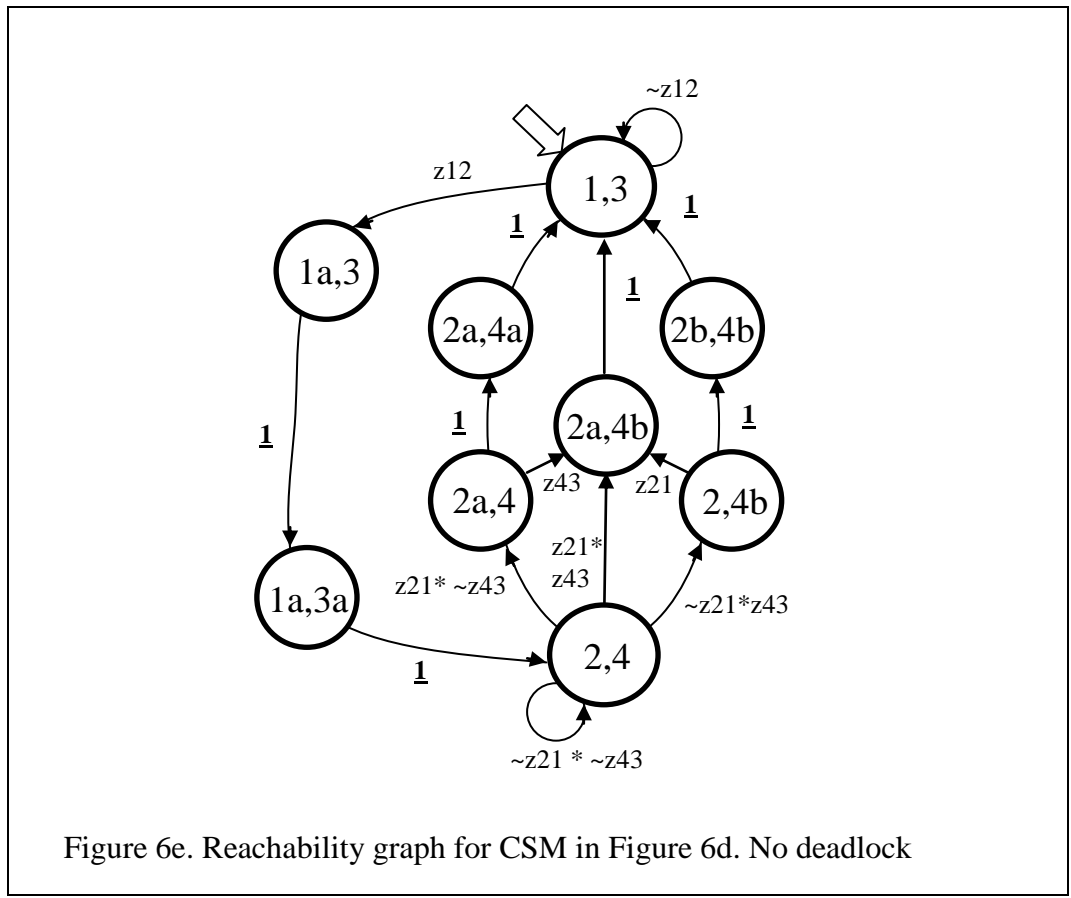

Figure 6e. Reachability graph for CSM in Figure 6d. No deadlock

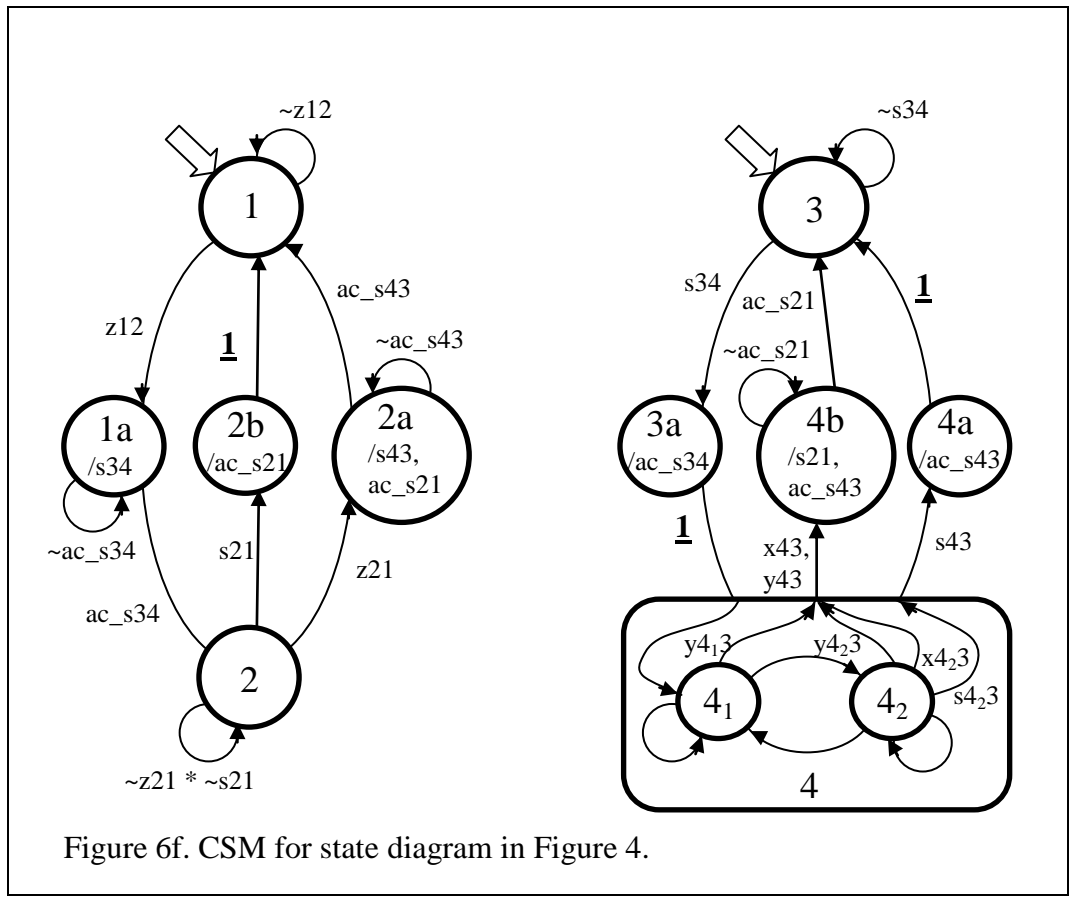

Figure 6f. CSM for state diagram in Figure 4.

## 4. ENFORCEMENT CONSTRAINTS IN ENGINEERING SOFTWARE

The engineering design could be, for simplicity, treated as searching for final values of group of parameters, called design parameters (DP). Nature of these parameters can be quite different: material, geometrical, electrical, architectural, etc. There are many types of engineering design. Some of those types are based on the iterative process of engineering [Basz95, Basz96a, Basz96b].

The engineering system design process should result in finding values of parameters fully describing the system. The design process should be preceded by the detailed analysis of physical phenomena outlighting detailed properties of the engineering system. Such analysis should lead to the creation of a physical model of the engineering system. The physical model includes the constraints on the design parameters (C_DP).

The application/market/utility requirements for the engineering system product result in some constraints on chosen properties of the designed system. In this paper, for simplicity, we assume that only one such property of the designed system is considered. We will refer to such property as output characteristics (OC). Constraints on output characteristics will be called C_OC. The criterion of evaluation of the quality of the output characteristics is based on comparison with the given ideal output characteristics (I_OC). To measure the closeness of these two characteristics an objective function (F_OF) will be used.

The designer must be given or must develop a mathematical model of the engineering system, giving complete mathematical relationships between all interesting quantities describing its internal and external behavior. The complete mathematical model also includes the precise algorithms of resolving of all mathematical formulas and equations describing the required relationships. Such mathematical model (F_OC) enables computation of the output characteristics for any set of values of design parameters.

It is a choice of the designer to use the proper design methods and accuracy. The choice affects the effectiveness of the design. Especially important may be planning of detailed strategy of searching for the final values of design parameters during design process. In many situations the best methods are based on an iterative process to find the final values of design parameters.

The system including two modules performing such engineering design is shown in Figure 7a. It is a simplified version, where all negative situations in iterative process (C_DP not proper, C_OC not held or F_OF much worse than optimum found) are represented by one event *err*. The event *d_r* stands for decision request signal, issued after any of enumerated negative situations. The shape of a system corresponds to that in Fig. 3. Numbers of states (excluding *START_DESIGN* and *END_DESIGN*) are exactly the same. Signal *err* corresponds to *y43*, while *si* corresponds to *s43*.

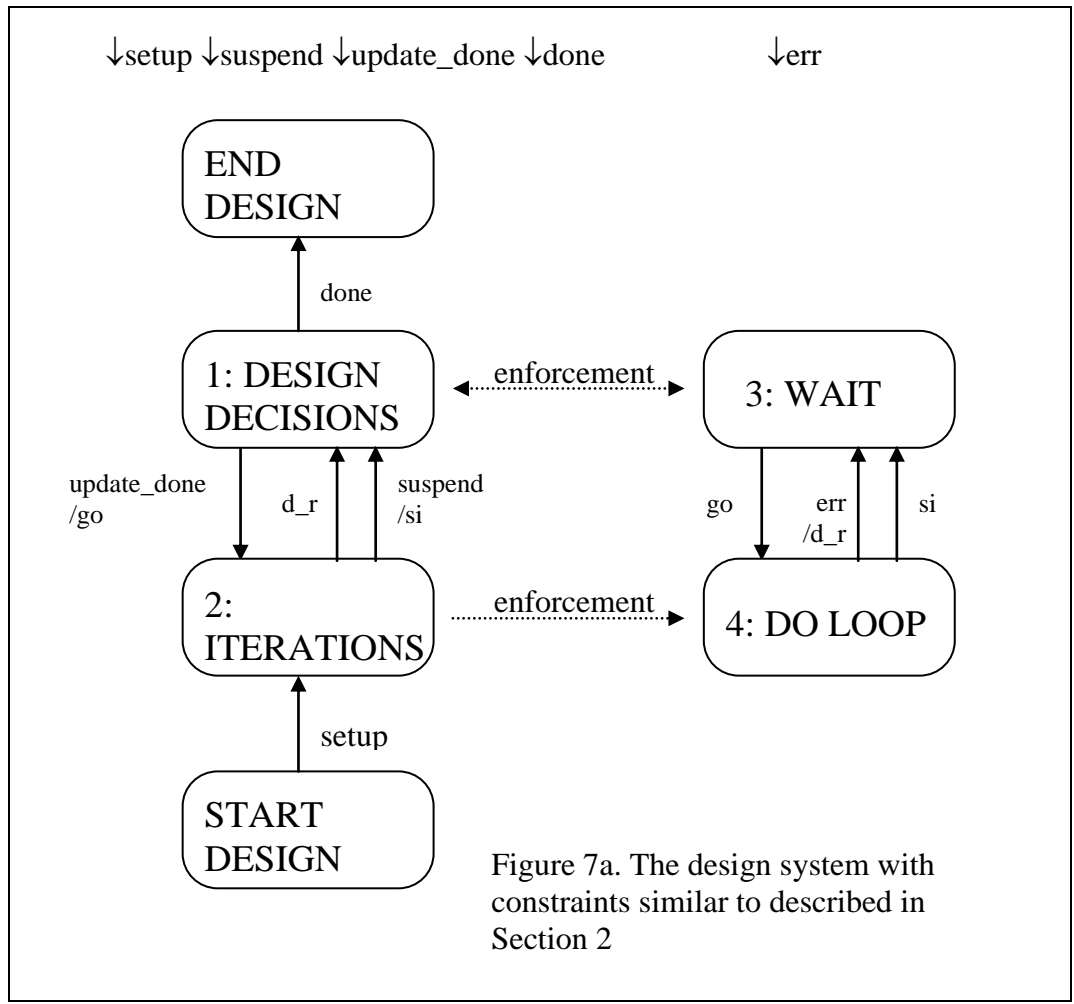

Figure 7a. The design system with constraints similar to described in Section 2

From this diagram with constraints we can generate automatically correct CSM similar to that in Figure 6c. The system now can be refined to reach a proper level of precision. An example of a refinement of a given state (DO_LOOP) is shown in Figure 7b. The state FIRST_CALCULATION corresponds to a state CC_DP_DECISION (preparation of input values for F_OF calculation). The state OTHER CALCULATIONS includes OC_EVALUATION, CC_OC_DECISION, OF_DECISION and DP_MODIFICATION states. A CSM specification generated form this diagram is similar to that in Figure 6f.

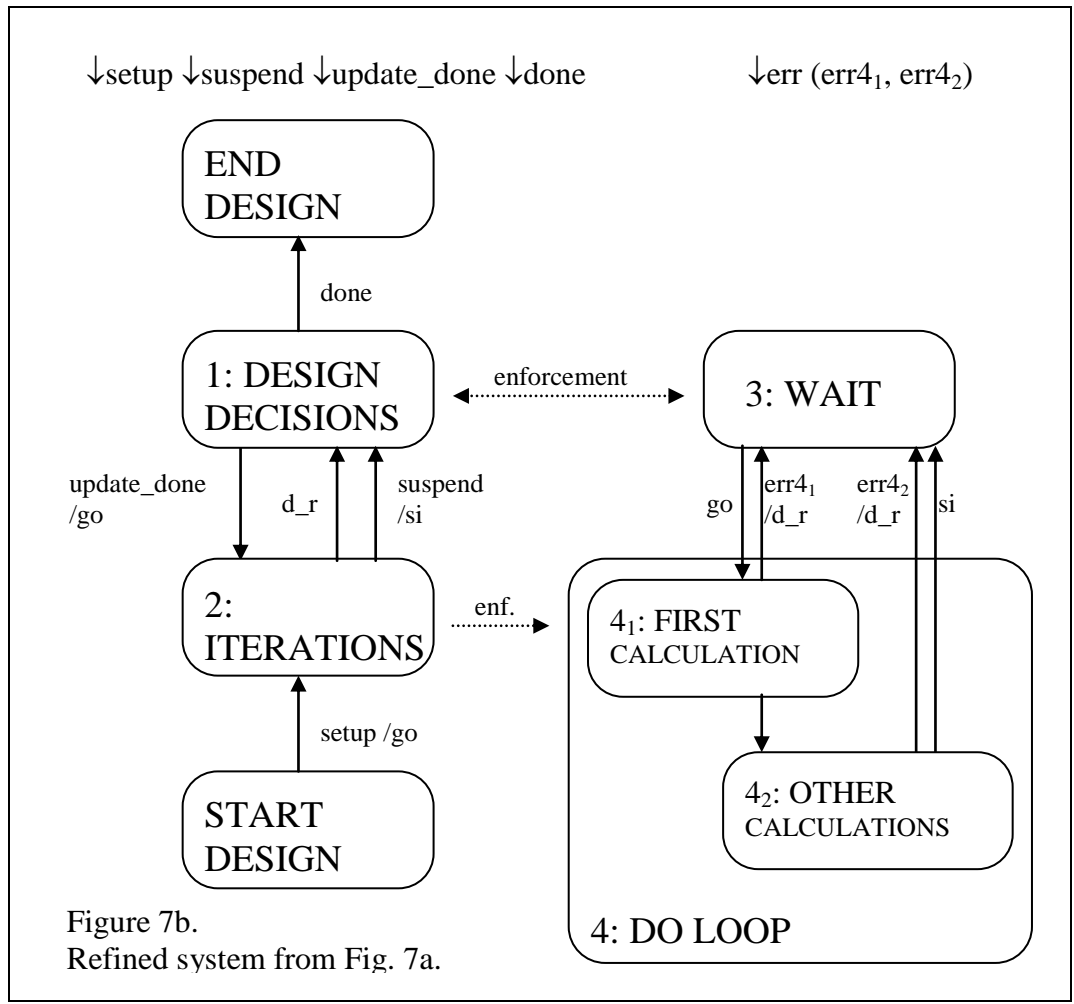

Figure 7b.
Refined system from Fig. 7a.

## 5. SUMMARY

In previous papers we have described how to create and verify concurrent software for engineering design. We have used the approach based on state diagrams software that is one of the most explicit and visual way of determining interactions between the software modules. However, as shown in our papers the synchronization of concurrent diagrams is a nontrivial issue and complete specification requires highly specialized knowledge. In this paper we have shown how to use the constraints on the state diagram to describe only the intentions of the engineer in respect to the interactions. We have also shown how the appropriate synchronization can be constructed automatically based on these constraints. We proposed generation of automata in CSM formalism. Semantics synchronization constraints can be checked statically by inspection of reachability graph of concurrent automata. The rules of inspection can be generated automatically by software that analyses synchronization constraints. Specialized software can generate a skeleton software of a system of concurrent programs implementing specific modules-automata. This software can additionally check if the constraints are held during execution.

In this paper we have concentrated on two-way enforcement constraints. However one-way constraints and other constraints such as preemption, exclusion etc. can be used similarly.